\documentclass[epj]{svjour}
\usepackage{graphics}
\usepackage{multirow}

\begin{document}
\title{Cosmogenic background study for a $^{100}$Mo-based bolometric demonstration experiment at China JinPing underground Laboratory}
\author{
W. Chen\inst{1} \and 
L. Ma\inst{1}\thanks{Email: malong@fudan.edu.cn} \and 
J. H. Chen\inst{1}\thanks{Email: chenjinhui@fudan.edu.cn} \and 
H. Z. Huang\inst{1,2} \and 
Y. G. Ma\inst{1}
}                     
\institute{Key Laboratory of Nuclear Physics and Ion-beam Application (MOE), Institute of Modern Physics, Fudan University, Shanghai, China \and Department of Physics and Astronomy, University of California, Los Angeles, California, USA}
%
%

\abstract{
We perform simulation study for a 10-kg $^{100}$Mo-based bolometeric demonstration experiment for neutrinoless double-beta decay ($0\nu\beta\beta$) search at China JinPing underground Laboratory (CJPL). Cosmogenic production of radionuclides in $^{100}$Mo-enriched lithium molybdate crystals and copper components of the detector system are studied using Geant4 toolkit based on the simulated cosmic ray data from the CRY library. Background energy spectra of the cosmogenic radionuclides including $^{56}$Co, $^{82}$Rb and $^{88}$Y which are harmful for the $^{100}$Mo-based $0\nu\beta\beta$ experiment are investigated. We then evaluate the total cosmogenic background level in the $^{100}$Mo $0\nu\beta\beta$ search energy region of interest (ROI) for the demonstration experiment. After one year of cooling down underground, the residual background contribution is found to be 1.8$\times$10$^{-6}$ cts/kg/keV/yr and 3.3$\times$10$^{-4}$ cts/kg/keV/yr from crystals and copper components, respectively. Furthermore, underground cosmogenic activation of copper and lithium molybdate crystal is calculated based on the simulation spectra of neutron and proton in CJPL. The underground cosmogenic background is found to be negligible in the ROI.
} 

\maketitle
\section{Introduction}
\label{intro}
Neutrinoless double-beta decay ($0\nu\beta\beta$) is important fundamental physics beyond the Standard Model. Without neutrino emitted, $0\nu\beta\beta$ decay violates the lepton number conservation~\cite{Bilenky:2012qi,DellOro:2016tmg}. This decay process can only take place if neutrinos are Majorana particle, i.e. neutrino is its own antiparticle~\cite{Furry:1939qr}. Experimental observation of the $0\nu\beta\beta$ decay will reveal the Majorana nature of neutrino and shed light on the studies of the absolute mass scale of neutrino and the orgin of the matter-antimatter asymmetry in our universe.

In the past decades, various technological approaches have been adopted for the experimental search of $0\nu\beta\beta$ decay worldwide. Though so far no evidence of the $0\nu\beta\beta$ decay is observed, experimental data have put stringent limits on the half-life of the $0\nu\beta\beta$ decay, which includes above 10$^{25}$ year for the double-beta decay nuclide $^{130}$Te and 10$^{26}$ year for $^{76}$Ge and $^{136}$Xe~\cite{GERDA:2020pmc,KamLAND-Zen:2016pfg,CUORE:2020bok,CUORE:2019yfd}.

Cryogenic crystal bolometer is favorable methodology among the $0\nu\beta\beta$ experimental approaches because of its good energy resolution, high detection efficiency and high long-term operational stability~\cite{Fiorini:1983yj}. The next-generation cryogenic crystal bolometeric experiment will be based on the CUPID technology (CUORE Upgrade with Particle IDentification) with expertise learned from the CUORE experiment~\cite{CUPID:2015yfg,CUPID:2019imh}. With the deployment of the scintillating bolometers and readouts of both heat and light channels, CUPID is able to do active background discrimination with the capability of eliminating the backgrounds from $\alpha$ particles which is a key technology of the future experiment to achieve substantial improvement in $0\nu\beta\beta$ discovery sensitivity.

Motivated by the successful development of the CUPID detector technology~\cite{CUPID:2015yfg,CUPID:2019imh}, in particular, the initial demonstration of the capability from CUPID-Mo result~\cite{CUPID:2020aow}, there is a strong scientific interest to develop a CUPID like scintillating crystal bolometer detector at China JinPing Laboratory (CJPL). The CUPID detector technology is cost effective, versatile and upgradeable with multiple cryogenic system to enhance sensitivity. A CUPID detector at CJPL could be a possible upgrade to the proposed CUPID experiment at the Laboratori Nazionali del Gran Sasso (LNGS), and contribute as part of the future ton-scale detector network envisioned by the CUPID collaboration~\cite{CUPID:2019imh}. The CJPL experiment will use lithium molybdate Li$_{2}$MoO$_{4}$ (LMO) crystal detector operating at extremely low temperature environment. CJPL, the deepest underground laboratory in the world with 2400 m rock overburden, provides ideal place for the experimental search of $0\nu\beta\beta$. The first phase of $^{100}$Mo-based bolometric experiment at CJPL will be operating a 10-kg demonstrator, which includes an array of LMO scintillating bolometers with heat-light dual readout to enable simultaneous detection of energy released as heat and light. 

For $0\nu\beta\beta$ rare event search, experiment needs to be operated in ultra-low background environment. Background level is one of the main limitations to the experimental sensitivity. To minimize the experimental background, continuous efforts have been made to understand the potential background sources and their contaminations to the signal region. For the bolometric experiment, environmental radioactivity, cosmic ray, radioimpurities in detector and shielding materials are significant sources that could contribute separately to the experimental background~\cite{Luqman:2016okt,CUORE:2017ztm}.  Among the background sources, radionuclides generated in materials from cosmogenic activation during the exposure to cosmic rays can introduce radioimpurity to the detector system which contribute considerable radioactive background to the experiment~\cite{Cebrian:2020bwn}. Although short-lived radioactive nuclides decrease rapidly, long-lived radionuclides can contribute background continuously to the energy spectrum in both low and high energy regions~\cite{Cebrian:2020bwn,Cebrian:2017oft,Barabanov:2005cw}. For the $^{100}$Mo-based bolometric demonstration experiment at CJPL, reliable evaluation of the background from cosmogenic activation is of great importance to control the background, which is essential for achieving the sensitivity goal of the demonstrator as well as future large scale experiment. 

In this paper, we investigate the cosmogenic activation in LMO crystals and copper components of the $^{100}$Mo-based bolometric demonstration experiment at CJPL by performing a Geant4 Monte Carlo simulation study. Background contributions to the $0\nu\beta\beta$ search ROI from cosmogenic radionuclides are evaluated. Cosmogenic production of the radionuclides and simulation setup for the demonstrator are discussed in Section 2. Results of the production rates, background energy spectra and quantitative evaluation of the background level are presented in Section 3. A brief summary is given in Section 4 . 

\section{Methodology}
\label{sec:1}

\subsection{Cosmogenic production of radionuclides}
\label{sec:1-1}

Cosmogenic radionuclides are produced by the interaction of cosmic ray particles with the target nucleus. The spallation of nuclei by high energy nucleons is one of the dominant processes that produces a series of radioactive unstable nuclides in the target material. These radionuclides decay accompanied by the release of gamma rays, electrons et al. particles, which will produce considerable background to the experiment. Cosmogenic activation studies have been performed in many underground experiments based on both experimental measurements and Monte Carlo simulation approaches~\cite{Juyal:2019gch,Ma:2018tdv,EDELWEISS:2016hgg,Back:2007kk,Mei:2009vn}. 

\begin{table}[htbp]
    \centering
\renewcommand{\arraystretch}{1.3}
    \begin{tabular}{ccc}
 \hline
        Isotope & Half-life($T_{\mathrm{1/2}}$) & Q-value(keV)\\
        \hline
         $^{22}$Na & 2.6 y & 2843\\
         $^{56}$Co & 77 d & 4567\\
         $^{60}$Co & 5.3 y & 2823\\
         $^{68}$Ga($^{68}$Ge) & 68 m (271 d) & 2921 (107)\\
         $^{82}$Rb($^{82}$Sr) & 1.27 m (25.4 d) & 4403 (178)\\
         $^{84}$Rb & 33 d & 2680\\
         $^{88}$Y($^{88}$Zr) & 107 d & 3623\\
 \hline
    \end{tabular}
    \caption{List of the potential harmful cosmogenic radioisotopes for $^{100}$Mo-based bolometric experiment selected by the Q-value ($>$ 2615 keV) and half-life ($>$ 20 days)}
    \label{table0}
\end{table}

Cosmogenic activation is highly related to the target material, cosmic ray flux and exposure time~\cite{Cebrian:2020bwn,Cebrian:2017oft,LAUBENSTEIN2009750}. Properties of the radionuclides including decay mode, decay rate, Q-value and the energy region of the experimental $0\nu\beta\beta$ search are all important factors need to be considered for evaluating the harmful cosmogenic background. For the CJPL $^{100}$Mo-based bolometric experiment, as short-lived nuclei can be eliminated simply by placing the detector system underground for a sufficiently long time before experimental data taking, therefore we consider those long-lived cosmogenic isotopes in crystals, detector tower frame and passive shielding components of the experiment. Radioisotopes with relative high production rate, long half-life, high decay Q-value and emission of high energy gamma-rays are selected and listed in Table~\ref{table0}. 

According to the $Q_{\beta\beta}$(~3034 keV) of $^{100}$Mo, the potential harmful cosmogenic radionuclide to the $^{100}$Mo-based bolometric experiment at CJPL are $^{56}$Co, $^{82}$Rb and $^{88}$Y produced in copper and Li$_{2}$MoO$_{4}$ crystals. The long-lived cosmogenic isotope $^{56}$Co exist commonly in the copper material. The characteristic gamma-rays emitted from these radioisotopes can reach the detector and yield significant background to the energy ROI of the $0\nu\beta\beta$ search. One should note that though the half-life of $^{82}Rb$ is short (1.25 minutes), its parent nuclide, the $^{82}$Sr has a relative long half-life about 25.4 days.  

In this study, we focus on the cosmogenic production of the long-lived radionuclides $^{56}$Co, $^{82}$Sr, $^{88}$Y, and $^{88}$Zr at sea level. The production rate of a nuclide can be calculated according to
\begin{equation}
    P = \sum_{i} N_i \sum_{j}\int_{0}^{\infty}\sigma_{ij}(E_{j})\phi_{j}(E_{j})dE_{j},
\label{q1}
\end{equation}
where $N_{i}$ is the number of nuclei of the target element $i$, $\sigma_{ij}(E)$ is the cross section for the production of a specific nuclide from the target element by cosmic ray particle $j$ with energy $E$, $\phi_{j}(E_{j})$ is the flux of cosmic ray particle $j$ (per unit energy). Different approaches can be adopted to calculate the production cross sections including COSMO, YIELDX, ACTIVIA packages~\cite{Back:2007kk,Silberberg:1998lxa,Martoff:1992oxa}. Other approaches like Monte Carlo (MC) simulations of the interaction between nucleons and nuclei are also widely used in cross section calculations.

Assuming a negligible level of the initial amount, the number of cosmogenic radionuclide produced in the target under cosmic ray exposure can be expressed by
\begin{equation}
    N(t_{expo}) = \frac{P}{\lambda}(1-e^{-\lambda t_{expo}}), 
\label{q2}
\end{equation}
where $P$ is the production rate of the cosmogenic radionuclide, $t_{expo}$ is the time of exposure to cosmic rays at sea level and $\lambda = ln2/T_{1/2}$ is the decay constant of the radionuclide. For long time exposure, there will be equal amount of nuclides produced and decayed, the production rate will become saturated. The number of radioisotopes after cooling underground for a time period can be expressed by
\begin{equation}
    N(t_{expo}+t_{cool}) = \frac{P}{\lambda}(1-e^{-\lambda t_{expo}})e^{-\lambda t_{cool}}, 
\label{q3}
\end{equation}
where $t_{cool}$ is the time of storage underground with shielding from cosmogenic activation. 

For some radionuclides, besides direct cosmogenic production, nuclear decay process can also contribute e.g.$^{88}$Y from $^{88}$Zr decay. In this case, the number of the nuclides can be expressed by
\begin{eqnarray}
    N(t_{expo}) = \frac{P_1 + P_2}{\lambda_1}(1-e^{-\lambda_1 t_{expo}})\nonumber\\
 + \frac{P_2}{\lambda_1 - \lambda_2}(e^{-\lambda_1 t_{expo}} - e^{-\lambda_2 t_{expo}}),
\label{q4}
\end{eqnarray}
where $P_1$ and $\lambda_1$ are the production rate and decay constant of the radionuclide, respectively. $P_2$ and $\lambda_2$ represent the production rate and decay constant of its parent nuclide. Taking into account both ground exposure and underground cooling, the number of radionuclides becomes
\begin{eqnarray}
    N(t_{expo}+t_{cool}) = \frac{P_1 + P_2}{\lambda_1}(1-e^{-\lambda_1 t_{expo}})e^{-\lambda_1 t_{cool}}\nonumber\\
 +\frac{P_2}{\lambda_1 - \lambda_2}(e^{-\lambda_1 t_{expo}} - e^{-\lambda_2 t_{expo}})e^{-\lambda_1 t_{cool}}\nonumber\\
 + \frac{P_2(1-e^{-\lambda_2 t_{expo}})(e^{-\lambda_1 t_{cool}} - e^{-\lambda_2 t_{cool}})}{\lambda_2 - \lambda_1}.
\label{q5}
\end{eqnarray}

\subsection{Monte Carlo simulation setup}
\label{sec:1-2}

Geant4 (version 4.10.05) is used to perform simulation study of the cosmogenic radionuclide production and their energy deposition in the detector. This simulation toolkit including a variety of particle-material interaction cross sections for describing the particle production and transport properties has been widely used for detector simulation~\cite{GEANT4:2002zbu,Allison:2016lfl}. Geant4 Monte Carlo simulation has been proven to be reliable for estimating cosmogenic nuclide production in many materials~\cite{Wei:2017hkq,Cebrian:2020bwn,LAUBENSTEIN2009750,Baudis:2015AP}. For the activation study, we use CRY (Cosmic-ray Shower Library) package to provide the information regarding energy, direction and flux of the cosmic ray particles~\cite{Papini:1996ai}. Cosmic muons, neutrons, protons, electrons, photons at sea level generated by the CRY package over a wide range of energies are used as inputs for the Geant4 simulation~\cite{4437209}. Physics list ``Shielding'' is selected for the simulating physics process. The ``Shielding'' physics list contains the electromagnetic and hadronic physics processes for high-energy and underground detector simulation. We have performed calculation from different physics list, the ``QGSP$\_$BERT$\_$HP" physics list with high precision neutron model implemented for neutrons at low energy, which was used in simulation studies for underground experiments~\cite{HWBae:2019AP}. The results are found to be compatible in trend and magnitude.

\begin{figure}[htbp]
    \centering
\resizebox{0.45\textwidth}{0.35\textwidth}{\includegraphics{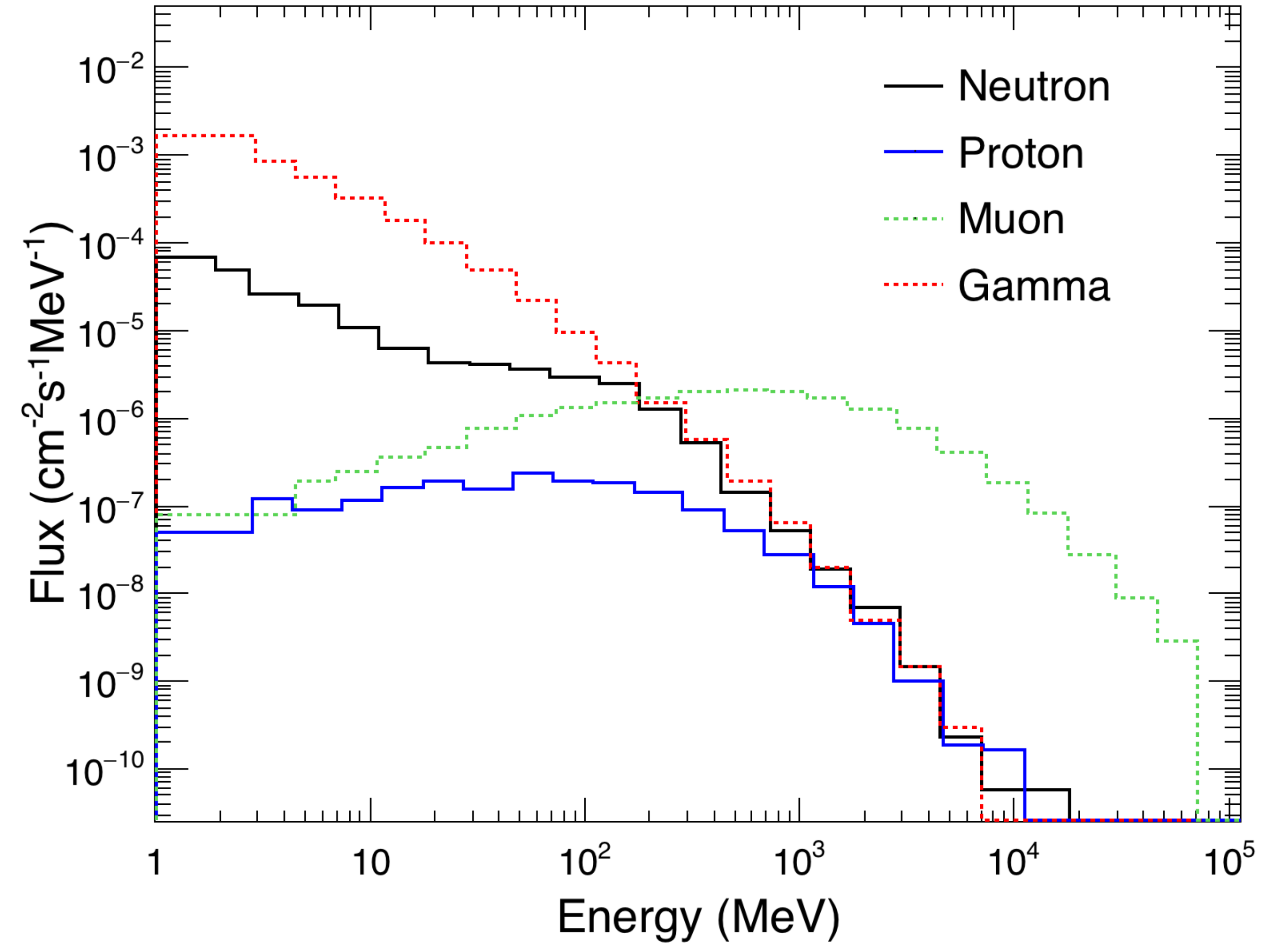}}
    \caption{(Color online) The energy spectra of cosmic ray particles at Shanghai sea level.}
    \label{fig:cosmic}
\end{figure}

\begin{table*}[htbp]
    \centering
\renewcommand{\arraystretch}{1.3}
    \begin{tabular}{c|cccc}
\hline
    Particle & Neutron & Proton & Gamma & Muon\\
    \hline
    Flux ($\rm cm^{-2}s^{-1}$) & $1.726_{-0.07}^{+0.136}\times10^{-3}$ & $1.386_{-0.046}^{+0.094}\times10^{-4}$ & $1.577_{-0.04}^{+0.08}\times10^{-2}$ & $1.134_{-0.027}^{+0.054}\times10^{-2}$\\
    \hline
    \end{tabular}
    \caption{The integral fluxes of cosmic-ray particles at Shanghai sea level. Uncertainties from the solar activity effect are calculated.}
    \label{table1}
\end{table*}

\begin{figure}[htbp]
    \centering
\resizebox{0.45\textwidth}{0.35\textwidth}{\includegraphics{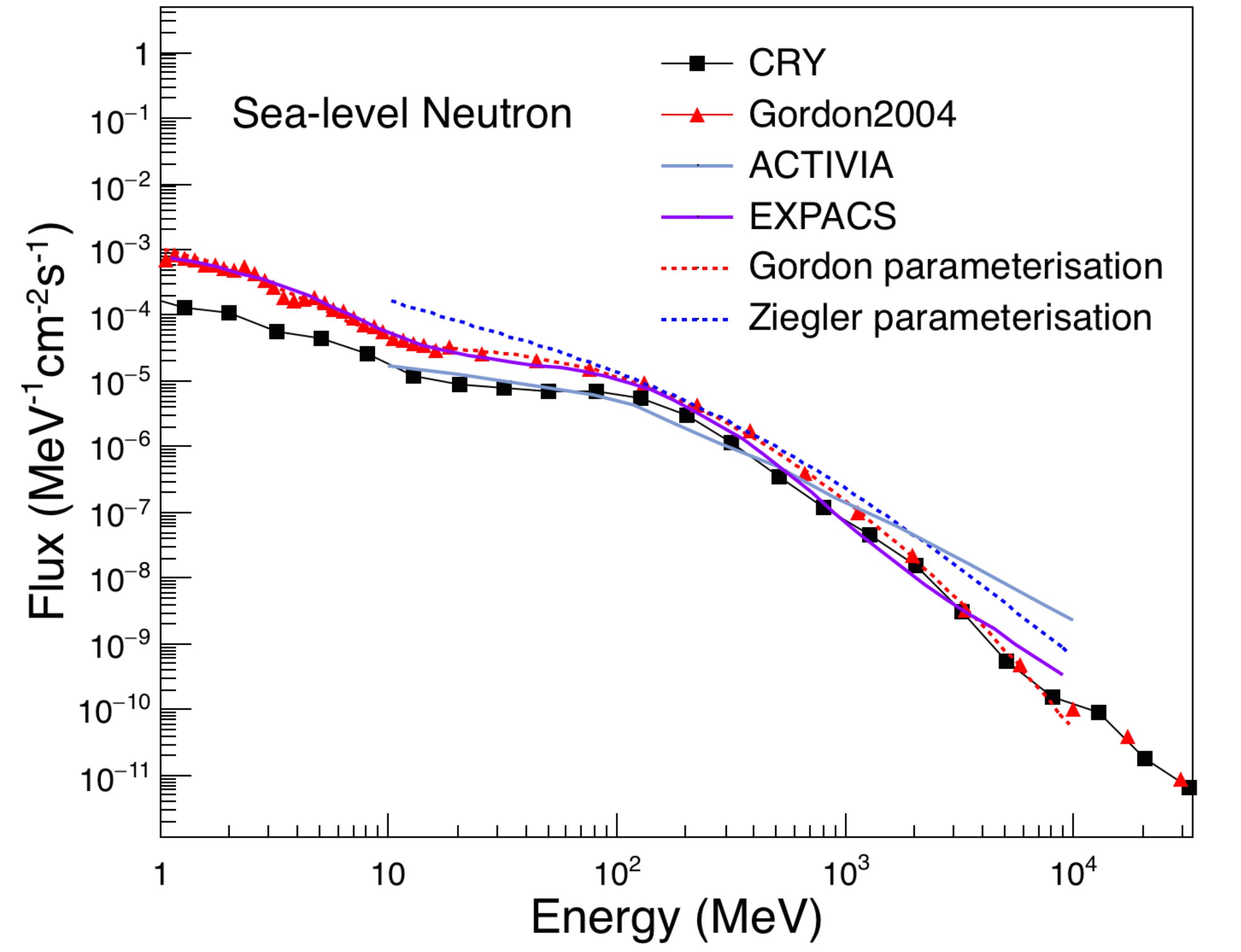}}
    \caption{(Color online) Comparison of the model and experimental measurement results of the cosmic ray neutron energy spectra.}
    \label{fig:neutron-spectra}
\end{figure}

Figure~\ref{fig:cosmic} shows the spectra of cosmic ray neutrons, protons, gammas, muons with energy above 1 MeV calculated using CRY. Integral fluxes are obtained for the various particles as shown in Table~\ref{table1}. The uncertainties shown represent the solar activity effect on the primary cosmic-rays. From the energy distribution, one sees that the differential flux spectra of neutron, proton and gamma are close above 1 GeV while muon flux is several orders of magnitude higher than other particles at high energy. One should note that the CRY results of the cosmic-ray spectra may differ from various model simulations as well as the experimental measurements. We further evaluate the uncertainties from the model descriptions of particularly the neutron and proton fluxes. Figure~\ref{fig:neutron-spectra} shows the comparison of model and experimental results of cosmic neutron spectra at New York sea level. Quantitative differences can be seen between the simulation results and data. Uncertainties of the flux modeling which will propagate to the cosmogenic-nuclide production rate is discussed in Section 3.1.

\begin{table}[htbp]
    \centering
\renewcommand{\arraystretch}{1.3}
    \begin{tabular}{cc}
 \hline
        Parameter & Value\\
        \hline
        Material & $^{100}$Mo enriched Li$_{2}$MoO$_{4}$\\
        Crystal size & 45$\times$45$\times$45 $\rm mm^{3}$\\
        Crystal mass & 0.28 kg\\
        Number of crystals & 36\\
        Total crystal mass & 10.08 kg\\
        $^{100}$Mo effective mass & 5.70 kg\\
 \hline
    \end{tabular}
    \caption{Main parameters of the $^{100}$Mo-based bolometric demonstration experiment at CJPL.}
    \label{table2}
\end{table}

Cosmogenic radionuclide production and background simulation are performed for the $^{100}$Mo-based bolometric demonstration experiment at CJPL. The key component of the demonstrator is an array of $^{100}$Mo enriched Li$_{2}$MoO$_{4}$ (LMO) crystals. The crystals are in the size of $45\times45\times45~\rm mm^3$ fixed to the copper frame by polytetrafluoroethylen (PTFE) supports. Increase of the temperature induced by the energy deposition due to particle interaction is read out through the Neutron-transmutation-doped (NTD) Ge thermistor. Thin germanium chips with NTD thermistor attached are used for scintillation light detection. The detector system works at extremely low temperature ($T\sim10$ mK) inside the cryogenic system. The $^{100}$Mo enriched lithium molybdate crystals are produced in the ground laboratory in Shanghai and transport to CJPL where the detector will be assembled. 

\begin{figure}[htbp]
    \centering
\resizebox{0.48\textwidth}{0.35\textwidth}{\includegraphics{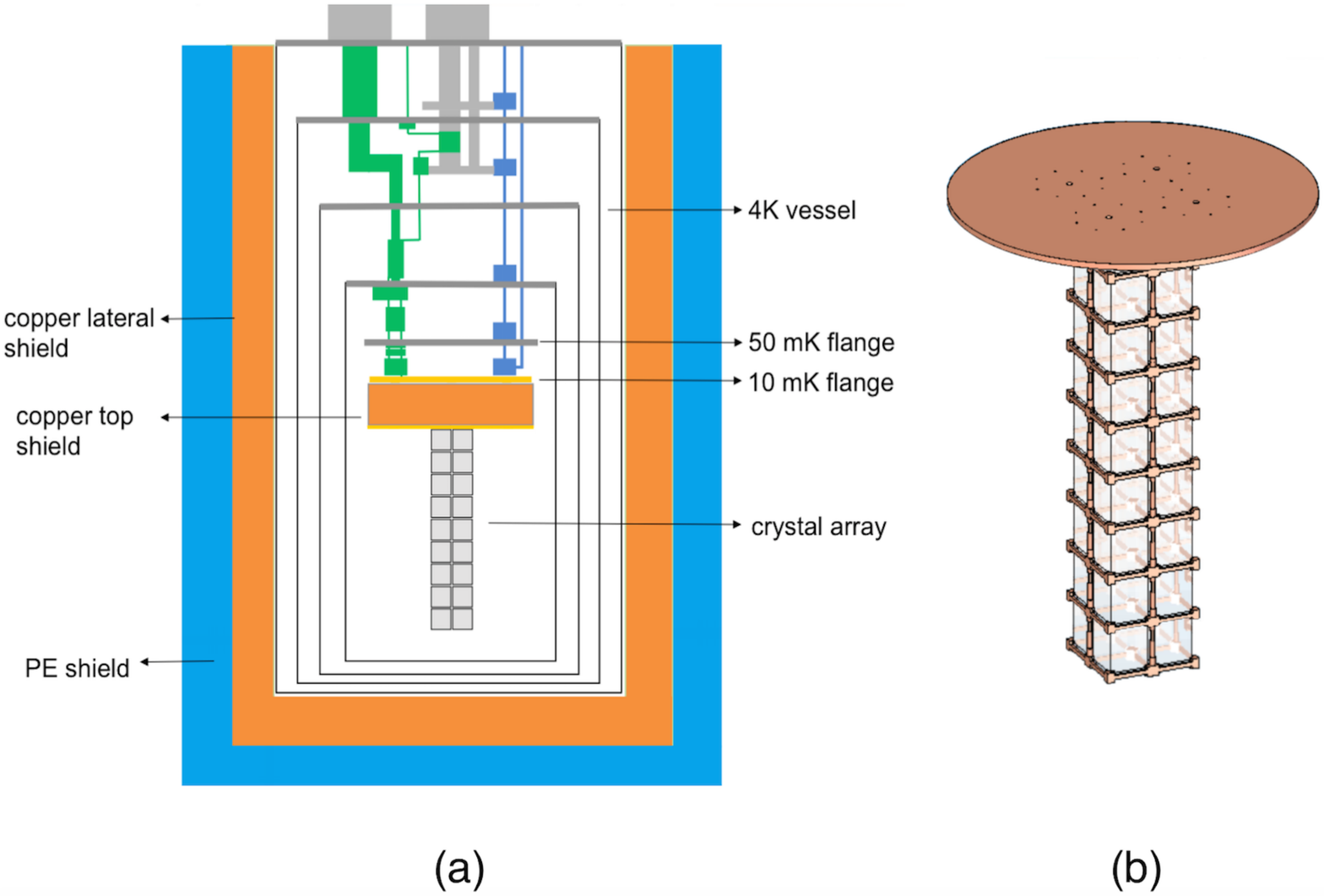}}
    \caption{Detector geometry implemented in the simulation study. (a) schematic view of the experimental setup including the cryogenic and shielding systems. (b) structure of the detector array.}
    \label{detectoranshield}
\end{figure}

The $^{100}$Mo-based bolometric demonstrator at CJPL aims at a background level below 10$^{-3}$ cts/keV/kg/yr in the $0\nu\beta\beta$ search energy ROI. To achieve the background goal, low radioactive materials are selected and used for the detector fabrication. The frame of the detector array and nearby shields are made of high purity oxygen-free electroforming copper (CuOFE) due to the low radioactivity, good thermal and mechanical properties of the material. The detector array is operated inside a customized cryogenic apparatus. Figure~\ref{detectoranshield} shows the schematic view of the detector and shielding configuration implemented in this simulation study. The cryostat is made up of three layers of cylindrical vessel acting as thermal shields of different temperature stages. The detector array is enclosed inside the vessels. In the cryostat, a 12 cm thick top copper shield is placed right above the detector tower to shield the detector from the radioactivity of the cryogenic system. Except the outer vacuum chamber, vessels are made of high purity CuOFE. The whole apparatus is a cryogen-free cryostat maintained by cooling system including pulse tube cryocoolers and $^3$He/$^4$He dilution refrigerator. The cryostat is surrounded by a 10 cm thick copper shield for external environmental gamma shielding and 15 cm thick borated polyethylene (PE) wall for neutron shielding. A bottom copper shield is placed below the cryostat to shield radioactivity from the laboratory floor. Table~\ref{table3} summarizes the main copper components and their weights.

\begin{table}[htbp]
    \centering
\renewcommand{\arraystretch}{1.3}
    \begin{tabular}{ccc}
\hline
        Component & Volume(cm$^{3}$) & Mass(kg)\\
        \hline
        Tower frame & 202.0 & 1.8\\
        Top copper plate & 1590.4 & 14.2\\
        Top copper shield & 23562.0 & 209.7\\
        50 mK thermal shield & 2224.7 & 19.8\\
        4 K thermal shield & 2404.5 & 21.4\\
        Lateral copper shield & 336000.0 & 2990.4\\
        Bottom copper shield & 30000.0 & 267.0\\
\hline
    \end{tabular}
    \caption{Copper components of the shielding system implemented in the simulation.}
    \label{table3}
\end{table}

\section{Results and discussions}
\label{sec:2}

\subsection{Production of the cosmogenic radionuclides in copper and LMO crystal}
\label{sec:2-1}

The key components of the detector in our experiment are $^{100}$Mo-enriched lithium molybdate crystals and high purity copper material used to build the detector array, thermal vessel and external background shields. Based on the Geant4 simulation, we focus on the study of the cosmogenic radionuclide production in copper and $^{100}$Mo-enriched lithium molybdate crystals. The results are scaled by the fluxes of cosmic neutron, proton, gamma, muon and normalized to cts/kg/day.

\begin{table*}[htbp]
    \centering
\renewcommand{\arraystretch}{1.3}
    \begin{tabular}{ccccccccc}
\hline
    Radionuclide & Target & Half-life(days) & Neutron & Proton & Gamma & Muon & Total & ACTIVIA\\
    \hline
    $^{56}$Co & Copper & 77.3 & 6.36 $\pm$ 0.51 & 0.85 $\pm$ 0.06 & 0.10 $\pm$ 0.01 & 0.02 $\pm$ 0.00  & 7.33 $\pm$ 0.51 & 8.70\\
    \hline
    $^{88}$Zr & LMO & 83.4 & 5.48 $\pm$ 0.43 & 0.77 $\pm$ 0.05 & 0.10 $\pm$ 0.01 & 0.07 $\pm$ 0.00 & 6.42 $\pm$ 0.43 & 5.18\\
    $^{88}$Y & LMO & 106.6 & 1.99 $\pm$ 0.15 & 0.28 $\pm$ 0.02 & 0.22 $\pm$ 0.01 & 0.00 & 2.47 $\pm$ 0.15 & 1.10\\
    $^{82}$Sr & LMO & 25.4 & 1.85 $\pm$ 0.16 & 0.33 $\pm$ 0.02 & 0.00 & 0.01 $\pm$ 0.00 & 2.19 $\pm$ 0.16 & 1.25\\
    $^{56}$Co & LMO & 77.3 & 0.04 $\pm$ 0.01 & 0.03 $\pm$ 0.00 & 0.00 & 0.00 & 0.07 $\pm$ 0.01 & 0.04\\
\hline
    \end{tabular}
    \caption{The production rates ($\rm kg^{-1}day^{-1}$) of cosmogenic radioisotopes induced by cosmic ray particles in LMO crystal and copper at Shanghai sea level. Uncertainties from the solar activity effect are also shown.}
    \label{table4}
\end{table*}

For the $^{100}$Mo-based experiment, as the Q-value of $^{100}$Mo $0\nu\beta\beta$ decay is above 3000 keV, the potential dangerous cosmogenic nuclides that could contribute background to the $^{100}$Mo $0\nu\beta\beta$ ROI are $^{56}$Co produced in copper and $^{56}$Co, $^{82}$Sr, $^{88}$Y, $^{88}$Zr produced in LMO crystals according to the characteristic gamma energy of the radionuclides. We simulate the production rates of $^{56}$Co, $^{82}$Sr, $^{88}$Y and $^{88}$Zr using Geant4. Table~\ref{table4} shows the simulation results of the production rates of cosmogenic radionuclides induced by different cosmic ray particles at Shanghai sea level. Since the half-life of $^{82}$Rb is short, production of $^{82}$Sr is studied instead of $^{82}$Rb. 

Besides the Geant4 simulation, different approaches can be applied to calculate the production cross section and evaluate the cosmogenic production rate~\cite{Silberberg:1998lxa,Martoff:1992oxa}. For comparison, we have calculated the production rate using ACTIVIA~\cite{Back:2007kk}, which is a popular computer package commonly used in calculation of the production and decay yields of isotopes from cosmic ray activation using data tables and semi-empirical formulae~\cite{Back:2007kk}. As ACTIVIA calculation of the radionuclide production in compound material is quite inefficient, to get the production rate of $^{82}$Sr or $^{88}$Y in the LMO crystal, we calculate the production rates in molybdenum first and then convert the results to that in LMO crystal according to:
\begin{equation}
    P_{LMO} = P_{Mo} \times f_{Mo}, 
\end{equation}
where $P_{\rm LMO}$ is the production rate of the radionuclide in lithium molybdate crystal, $f_{\rm Mo}$ is the relative spatial density of Mo in the LMO crystal, with $f_{\rm Mo}$ $\approx$ 0.56 for a 95$\%$ $^{100}$Mo enrichment.

\begin{table*}[htbp]
    \centering
\renewcommand{\arraystretch}{1.3}
    \begin{tabular}{ccccccc}
\hline
    Radionuclide & Target & CRY & Gordon2004 & ACTIVIA & Ziegler & EXPACS\\
    \hline
    $^{56}$Co & Copper & 8.32 & 15.75 & 6.30 & 14.47 & 11.98\\
    \hline
    $^{88}$Zr & LMO & 7.15 & 14.16 & 5.72 & 14.30 & 11.02\\
    $^{82}$Sr & LMO & 2.39 & 4.54 & 2.10 & 4.19 & 3.00\\
    $^{56}$Co & LMO & 0.05 & 0.08 & 0.15 & 0.13 & 0.04\\
\hline
    \end{tabular}
    \caption{Cosmic-ray neutron induced production rates ($\rm kg^{-1}day^{-1}$) of the cosmogenic radioisotopes at New York sea level.}
    \label{table5}
\end{table*}

For the activiation study, one million events are generated for each kind of the cosmic ray particle. Results of the production rate estimated by Geant4 are summarized in Table~\ref{table4}, where a comparison with ACTIVIA is presented as well. Reasonable agreement between the Geant4 and ACTIVIA results is found considering the differences in the input spectra of the cosmic ray particles and cross-section databases. We find that cosmic ray neutron induced interactions dominate the cosmogenic radionuclide production at Shanghai sea level. A majority of the cosmic ray protons are supposed to be absorbed by the atmosphere. Neutron-induced $^{56}$Co in copper or $^{88}$Y in LMO crystal are found to contribute approximately 80$\%$-85$\%$ of their total cosmogenic yield. Cosmogenic activation of copper has been extensively studied through Monte Carlo simulation. Quantitative differences can be found in the cosmic ray flux and model descriptions of the beam energy dependence of the isotope production cross section. For the cosmic-ray flux simulation, besides CRY model, other models e.g. EXPAC, Gordon and Ziegler parameterization were also used in previous studies~\cite{Cebrian:2020bwn,EXPACS,Gordon2004}. We evaluate in particular the uncertainties of our Geant4-based simulation by comparing the results of the production rate of $^{56}$Co, $^{88}$Zr and $^{82}$Sr from different model calculations. It is found both cosmic ray spectrum and production cross section contribute to the uncertainty of the radioisotope production rate. Table~\ref{table5} summarizes the Geant4 results of the neutron-induced production rates based on different model inputs of the New York cosmic-ray spectra. The relative differences are found within 15-90$\%$. Similarly, we estimate the uncertainties from the production cross section predictions by comparing Geant4 results with ACTIVIA, Talys and Silberberg-Tsao parameterization. The relative uncertainties are found not exceeding 85$\%$. With scaling to the same integral flux, the production rate of $^{56}$Co in our study is consistent within 20$\%$ in magnitude with previous studies~\cite{Ma:2018tdv,Barabanov:2005cw}.

\subsection{Cosmogenic background study for the $^{100}$Mo-based bolometric demonstration experiment at CJPL}
\label{sec:2-2}

Based on the production rates of the radionuclides, we evaluate the cosmogenic background for the $^{100}$Mo-based bolometric demonstration experiment at CJPL. To simulate the background distribution, we generate cosmogenic radionuclides uniformly in crystals and copper components of the simulation setup as shown in Fig.~\ref{detectoranshield}. The energy deposition of the radioactive decays in the detector is recorded for energy spectrum reconstruction.

\begin{table}[htbp]
\centering
\renewcommand{\arraystretch}{1.3}
\begin{tabular}{ccc}
\hline
Material & Process & Duration (days) \\ 
\hline
\multirow{2}*{LMO crystal} & Growth & 30 \\
~ & Machining/Tansportation & 15 \\
\hline
\multirow{2}*{CuOFE} & Production & 60 \\
~ & Machining/Tansportation & 15 \\
\hline
\end{tabular}
\caption{Estimated time consumption of the main processes of the material preparation before shipping to CJPL.}
\label{table6}
\end{table}

According to the average time consumption of the LMO crystal production including crystal growth, machining and transportation and similarly the time cost in ground preparation of the copper components, in this study, a total exposure time of 45 days and 75 days are assumed for LMO crystal and copper, respectively. The main processes and estimated time of the crystal and copper production are listed in Table~\ref{table6}. Though most of the cosmogenic isotopes could be eliminated by zone refinement and crystallization in the crystal growth with rigorously controlled processing procedures, we consider cosmogenic activation during the whole crystal production process in current study.

\begin{table*}[htbp]
    \centering
\renewcommand{\arraystretch}{1.3}
    \begin{tabular}{cccccc}
\hline
\multirow{2}{*}{Radionuclide} & \multirow{2}{*}{Target} & \multicolumn{4}{c}{Cooling time (days)} \\ \cline{3-6}
    ~&~& 90 & 180 & 360 & 720\\
    \hline
    $^{56}$Co &  Copper & 151.8 $\pm$ 12.1 & 67.7 $\pm$ 5.4 & 13.5 $\pm$ 1.1 & $(5.0 \pm 0.4)\times10^{-1}$\\
    $^{88}$Zr & LMO & 80.7 $\pm$ 6.0 & 38.2 $\pm$ 2.9 & 8.5 $\pm$ 0.6 & $(4.2 \pm 0.3)\times10^{-1}$\\
    $^{88}$Y  & LMO & 114.2 $\pm$ 8.5 & 94.6 $\pm$ 7.1 & 44.5 $\pm$ 3.3 & 6.1 $\pm$ 0.5 \\
    $^{82}$Sr & LMO & 3.8 $\pm$ 0.3 & $(3.2 \pm 0.2)\times10^{-1}$ & $(2.4 \pm 0.2)\times10^{-3}$ & $(1.3 \pm 0.1)\times10^{-7}$\\
    $^{56}$Co & LMO & $(8.3 \pm 0.7)\times10^{-1}$ & $(3.9 \pm 0.3)\times10^{-1}$ & $(7.3 \pm 0.6)\times10^{-2}$ & $(2.9 \pm 0.2)\times10^{-3}$\\
\hline
    \end{tabular}
    \caption{The residual amount of radionuclides (in unit of cts/kg) in LMO crystals and copper shields after cooling underground for different time periods. Uncertainties from the solar activity effect on cosmic ray flux are shown.}
    \label{table7}
\end{table*}

The amount of long-lived cosmogenic radionuclides decreases over time. For cosmogenic background study, we estimate the residual amount of the cosmogenic radionuclides in the material for different cooling times, as shown in Table~\ref{table7}. We find that the number of $^{56}$Co, $^{82}$Sr and $^{88}$Zr per kilogram decreases by more than one order of magnitude after 360 days of cooling underground. For the relative short half-life of $^{82}$Sr, its residual amount decreases significantly by increasing the underground cooling time.

Considering the Q-value of the $0\nu\beta\beta$ decay of $^{100}$Mo and the energy resolution of the detector, we focus on the $^{100}$Mo $0\nu\beta\beta$ ROI (3010-3060 keV) in cosmogenic background evaluation. The energy resolution in our simulation study is modeled with
\begin{equation}
\sigma(E)=\sqrt{0.49+(0.058 \times \sqrt{E})^{2})},
\label{q7}
\end{equation}
according to the detector calibration results of the CUPID-Mo experiment~\cite{CUPID:2019vfp}.

\begin{figure}[htbp]
    \centering
\resizebox{0.45\textwidth}{0.35\textwidth}{\includegraphics{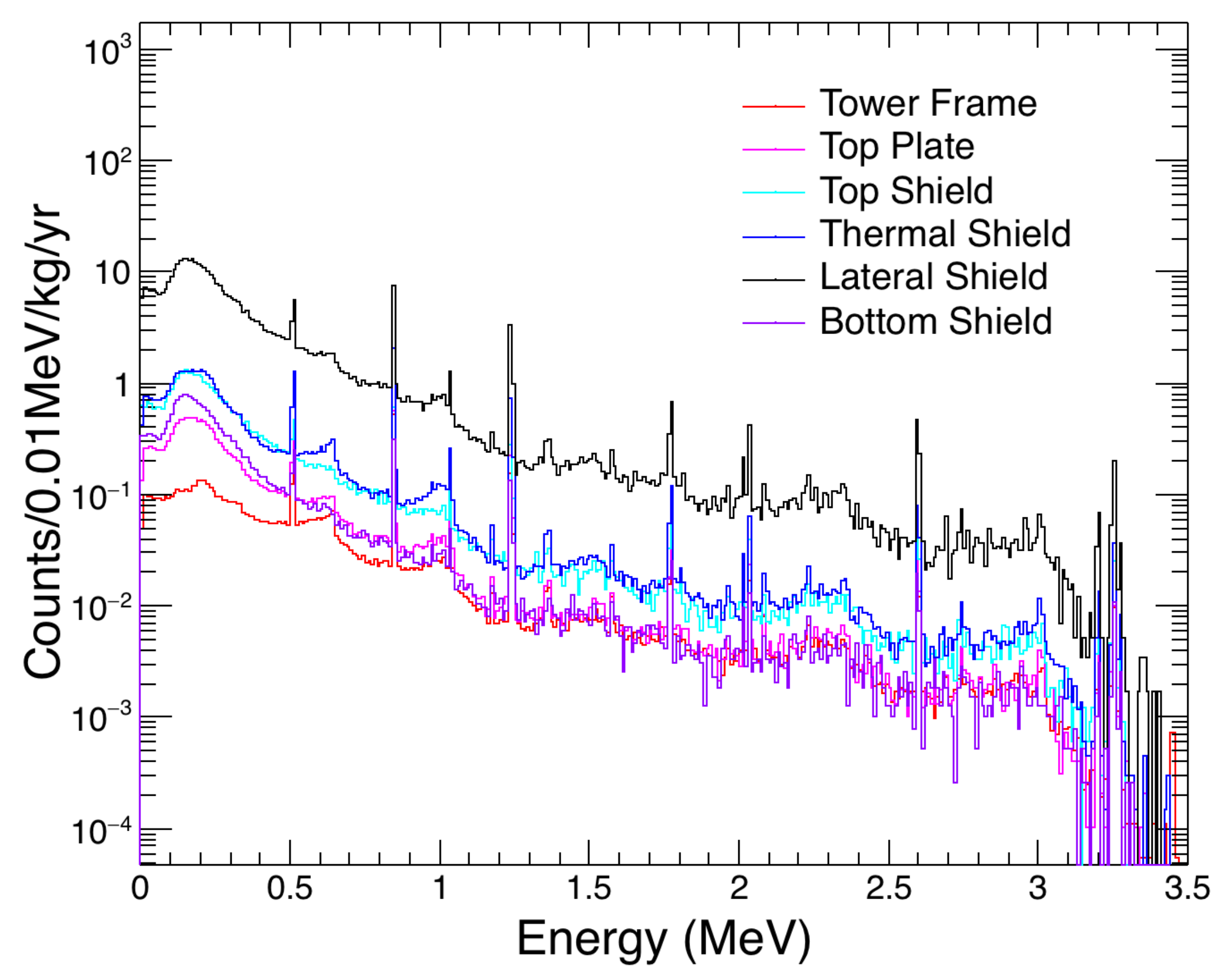}}
    \caption{(Color online) Simulated energy spectra for $^{56}$Co in different copper components of the experimental setup. The results are shown in different colors.}
    \label{fig4}
\end{figure}

Based on the production rates and decay properties of the radionuclides, we obtain the background energy spectra for different cosmogenic radionuclides of the $^{100}$Mo-based bolometric demonstration experiment at CJPL. Figure~\ref{fig4} shows the $^{56}$Co spectra results of the main copper components in the experiment. For all energy spectra, characteristic gamma peaks can be seen over the whole energy region. It is found that the $^{56}$Co background is mostly from the lateral shield outside of the cryogenic system. The massive material contributes continuous background to the ROI which is nearly an orders of magnitude higher than those from other copper components. 
 
\begin{figure}[htbp]
    \centering
\resizebox{0.45\textwidth}{0.35\textwidth}{\includegraphics{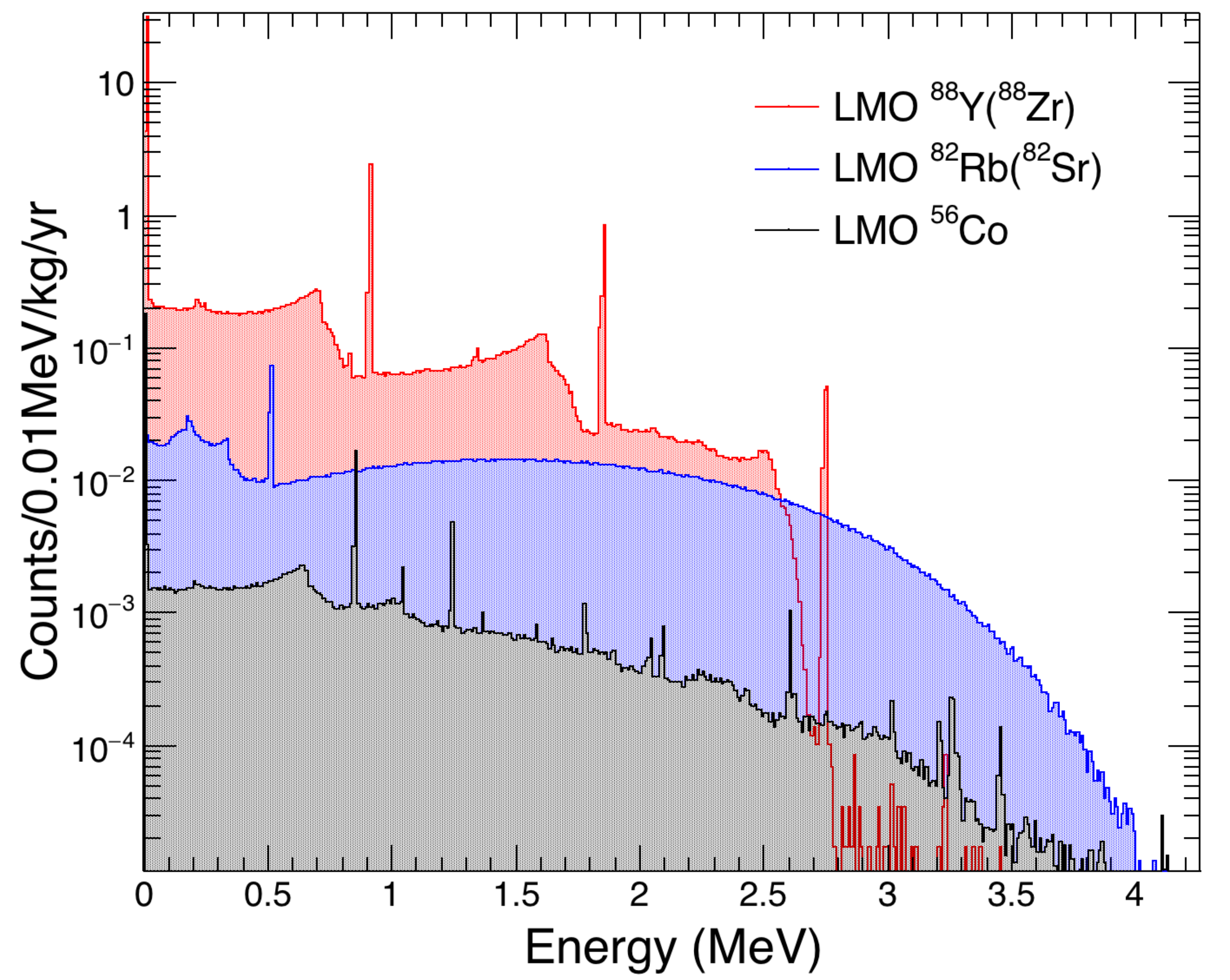}}
    \caption{(Color online) Simulated energy spectra of $^{56}$Co, $^{88}$Y and $^{82}$Rb ($^{82}$Sr) in the LMO detector with a 90 days cooling time.}
    \label{fig5}
\end{figure}

Similarly, we evaluate the background contributions from $^{88}$Y, $^{82}$Rb and $^{56}$Co produced in the LMO crystals. For 45 days of ground production and 90 days of underground cooling, the background energy spectra of these radioisotopes are shown in the  Fig.~\ref{fig5}. We find that $^{82}$Rb ($^{82}$Sr) and $^{56}$Co backgrounds extend to high energy region and contribute significantly to the ROI despite their low cosmogenic production rates.

\begin{figure}[htbp]
    \centering
\resizebox{0.45\textwidth}{0.35\textwidth}{\includegraphics{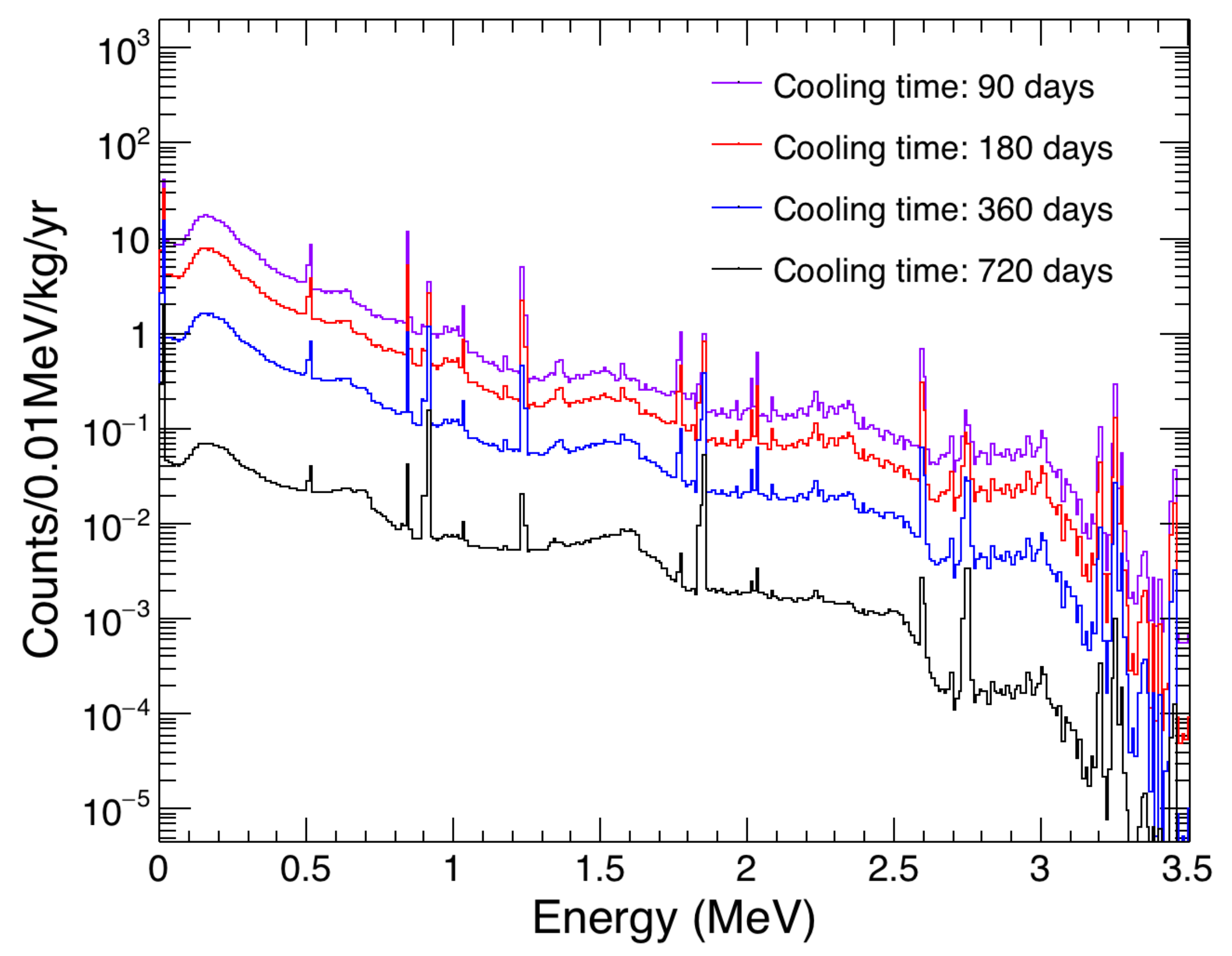}}
    \caption{(Color online) The total cosmogenic background energy spectra of the radioisotopes for different cooling times.}
    \label{fig6}
\end{figure}

\begin{table*}[htbp]
    \centering
\renewcommand{\arraystretch}{1.3}
    \begin{tabular}{ccccc}
\hline
\multirow{2}{*}{Component} & \multicolumn{4}{c}{Cooling time (days)}\\ \cline{2-5}
        ~ & 90 & 180 & 360 & 720\\
        \hline
        Copper & $(3.71\pm0.26)\times10^{-3}$ & $(1.65\pm0.11)\times10^{-3}$ & $(3.30\pm0.23)\times10^{-4}$ & $(1.27\pm0.09)\times10^{-5}$\\
        LMO & $(2.88\pm0.20)\times10^{-4}$ & $(2.78\pm0.19)\times10^{-5}$ & $(1.83\pm0.13)\times10^{-6}$ & $(1.42\pm0.10)\times10^{-7}$\\
        \hline
        Total & $(3.99\pm0.26)\times10^{-3}$ & $(1.68\pm0.11)\times10^{-3}$ & $(3.32\pm0.23)\times10^{-4}$ & $(1.28\pm0.09)\times10^{-5}$\\
\hline
    \end{tabular}
    \caption{The background level (in unit of cts/kg/keV/yr) in ROI for different cooling times. Uncertainties from the cosmic ray flux are shown.}
    \label{table8}
\end{table*}

Figure~\ref{fig6} shows the total cosmogenic background distributions for different cooling times at CJPL. The background level in the $^{100}$Mo $0\nu\beta\beta$ ROI is summarized in Table~\ref{table8}. One can find that the cosmogenic background is dominated by the copper components. After cooling for one year underground, the residual background from both copper and LMO are reduced by more than one order of magnitude. By extending the underground storage time, the total background rate in the ROI is expected to be lower than $10^{-4}$ cts/kg/keV/yr. 

Considering the uncertainties of the production rates of the cosmogenic isotopes, by convoluting the relative uncertainties of both cosmic ray flux and production cross section, the final background is found to be in the level of $(1.9-7.2)\times10^{-4}$ cts/keV/kg/yr for copper components and $(1.3-3.9)\times10^{-6}$ cts/keV/kg/yr for LMO crystals. 

One should note that passive shielding is not considered in our study for the ground production, neither for the transportation of crystal and copper. Appropriate cosmic ray shielding during ground transportation is expected to greatly reduce the cosmogenic activation of materials from cosmic ray exposure and thus reduce background from cosmogenic radionuclides~\cite{Ma:2018tdv}. 

\subsection{Evaluation of the underground cosmogenic background}
\label{sec:2-3}

In deep underground laboratory, the cosmic ray flux is significantly suppressed due to the thick rock overburden. The muons flux in CJPL is measured to be nearly eight orders of magnitude lower than that on the Earth's surface~\cite{Zhang:2016rlz}. Given the low level of the underground cosmic ray flux, the cosmogenic activation in CJPL is also supposed to be tiny~\cite{Wu:2013cno}. 

We estimate the cosmogenic activation and cosmogenic radioactive background contribution to the $^{100}$Mo-based bolometric demonstration experiment at CJPL from several important radioisotopes by simulation study. Though underground cosmic ray muons are negligible, muon-induced neutrons and protons could become important sources of the cosmogenic background which are harmful for the underground experimental search for $0\nu\beta\beta$ rare event.

\begin{table}[htbp]
    \centering
\renewcommand{\arraystretch}{1.3}
    \begin{tabular}{|c|c|c|}
\hline
        Material & Radionuclide & Saturated yield (cts/kg)\\
        \hline
        CuOFE & $^{56}$Co & 8.5x$10^{-6}$\\
        \hline
        \multirow{3}*{LMO crystal} & $^{88}$Y & 1.5x$10^{-5}$\\
        ~ & $^{88}$Zr & 8.4x$10^{-6}$\\
\hline
    \end{tabular}
    \caption{The saturated yields of the underground cosmogenic radionuclides produced by cosmic ray induced neutrons and protons at CJPL.}
    \label{table9}
\end{table}

Based on the simulation results of the neutron and proton spectra at CJPL, production of the radionuclides in copper and LMO crystal from underground cosmogenic activation are studied by Geant4~\cite{Zeng:2020cyw}. For long-time experimental operation underground, the production rates of the cosmogenic nuclides in the detector material are expected to reach saturation. The saturated yields of $^{56}$Co in copper and $^{88}$Y ($^{88}$Zr) are summarized in Table~\ref{table9} ($^{82}$Sr and $^{56}$Co yields are much smaller and thus can be neglected). Besides, we examined short-lived cosmogenic isotopes produced in the LMO crystal with high Q-value above the $0\nu\beta\beta$ threshold and copper with high energy gamma emissions. The potential radioisotopes with relative high production probability (e.g. $^{62}$Cu, $^{57}$Ni) are studied. We find that the production rates of the underground cosmogenic radionuclides in our study are extremely low. The total cosmogenic background from the underground radioisotopes in copper and LMO crystals is estimated to be in the level of 10$^{-10}$ cts/kg/keV/yr in the ROI which is several orders of magnitude lower than that of the cosmogenic radioisotopes from ground exposure. Further simulation study of the neutron-induced background in CJPL based on experimental measurements of the CJPL in-situ neutron spectrum is necessary for a comprehensive understanding of the underground cosmogenic background.

\section{Conclusion}
\label{sec:3}

We present systematic simulation study of the cosmogenic background from radioactive nuclides for a 10-kg $^{100}$Mo-based bolometric demonstration experiment for the neutrinoless double-beta decay search at China JinPing underground Laboratory. Cosmogenic activation in the LMO crystals and copper components including tower frame and copper shields are investigated using Geant4 software. The production rates of $^{56}$Co in copper and $^{82}$Rb ($^{82}$Sr), $^{88}$Y ($^{88}$Zr), $^{56}$Co in lithium molybdate (LMO) crystals at sea level are obtained based on the cosmic neutron, proton, gamma and muon pectrum from CRY library. Comparisons are made with the ACTIVIA calculations. It is found that cosmogenic production of radionuclides in the LMO crystal is dominated by cosmic neutron induced interactions similar as that in copper material. Based on the production rates of the radionuclides, energy spectra of cosmogenic $^{56}$Co, $^{82}$Rb and $^{88}$Y in both copper and LMO crystals are simulated and their background contributions to the ROI are evaluated. We compare the energy spectra for different underground cooling times. A total background level of $3.3\times10^{-4}$ cts/keV/kg/yr in the ROI is predicted for one year's underground cooling down based on our Geant4$+$CRY simulation study. The residual background is found to be dominated by the copper components especially the external copper shields.

For $0\nu\beta\beta$ bolometric experiment, cosmogenic radionuclides in detector system contribute non-negligible radioactive background. Experimentally, multiple approaches can be adopted to mitigate the cosmogenic background. As radionuclide activity decreases with increasing underground cooling time, by minimizing the exposure time above ground and maximizing the storage time underground, the cosmogenic activation in the experiment is expected to be effectively reduced to a quite low level. The future $^{100}$Mo-based bolometric experiment at CJPL aims at a total background goal of 10$^{-4}$ cts/keV/kg/yr. Besides long time underground cooling before experimental operation, underground copper production, machining and detector fabrication are all necessary for future experiment to achieve the background goal.

\emph{Acknowledgements}
This work is supported in part by the National Natural Science Foundation of China under contract Nos. 12025501, 11890710, 11890714, 12147114 and by the Strategic Priority Research Program of Chinese Academy of Sciences with Grant No. XDB34030200.

%
\bibliographystyle{unsrt}
\bibliography{draftNotes}
%
%
%

\end{document}